\documentclass{article}
\usepackage{spconf,amsmath,graphicx,hyperref}
\usepackage{booktabs}
\usepackage{multirow}

\title{Evaluating perceptual transparency of \\lossy music audio compression with Opus codec}
\name{Benjamin Martin, Romain Hennequin}
\address{Deezer Research\\
\texttt{research@deezer.com}}
\begin{document}
%\ninept
%
\maketitle
\begin{abstract}
Although the open-source Opus audio codec is widely used for digital audio streaming, its perceptual transparency threshold for high-bitrate music compression remains insufficiently evaluated. To establish this threshold, this study evaluates Opus across a two-phase ITU-R BS.1116-3 double-blind framework, analyzing Difference Grades using a Linear Mixed-Effects Model (LMM) and Two One-Sided Tests (TOST) for statistical equivalence. In a first phase with 21 qualified listeners and 23 music excerpts, 128~kbps achieves statistical in-distinguishability ($p = 0.456$) but narrowly fails strict equivalence to the uncompressed reference ($p = 0.064$). In Phase 2, Best Linear Unbiased Predictors (BLUPs) isolate the 10 most critical listeners and 8 most challenging samples for a stress-test. Under these worst-case conditions, 128~kbps is conclusively detected as degraded ($p = 0.030$), whereas 160~kbps successfully reaches perceptual transparency ($p = 0.747$) and strict statistical equivalence ($p = 0.024$). The study concludes that 160~kbps demonstrates perceptual transparency to uncompressed CD quality in critical listening environments. For reproducibility, we publicly release results for both phases, anonymized subject background info and sample characteristics~\footnote{All results available at https://zenodo.org/records/22830003}.
\end{abstract}
%

% \begin{figure}
%     \centering
%     \includegraphics[width=\linewidth]{figs/perceptual_scale.png}
%     \caption{Trial interface, showing the 3 synchronized audio players, seek bar, and ITU-R subjective scale.}
%     \label{fig:trial}
% \end{figure}

\section{Introduction}
\label{sec:intro}

Perceptual audio coding algorithms constitute the primary mechanism for the distribution and streaming of digital music across modern telecommunication networks. Their high level of efficiency is realized by exploiting the physiological and cognitive limitations of the human auditory system.

Standardized by the Internet Engineering Task Force (IETF) in RFC 6716~\cite{valin2012definition}, the Opus codec has emerged as a highly versatile and efficient open-source audio format for both interactive and streaming applications. Its hybrid architecture integrates the technology of the SILK codec, optimized for human speech, with the CELT codec, designed for high-quality music. This combination allows Opus to scale from low-bitrate narrowband speech to high-fidelity fullband stereo music.
Opus achieves superior compression by combining implicit bit allocation, Pyramid Vector Quantization (PVQ) for spectral preservation, and advanced psychoacoustic techniques, allowing it to outperform legacy codecs like Vorbis, MP3, and HE-AAC at equivalent bitrates~\cite{valin2016high}.

Current empirical evaluations of the Opus codec focus on speech, leaving a significant gap regarding high-bitrate music compression. To address this, our study investigates the core research question: what is the minimum bitrate required for Opus to achieve perceptual transparency and become subjectively indistinguishable from a lossless CD quality reference?

\section{Related work}
\label{sec:related}

Most existing literature on the perception of the Opus audio codec focuses on lower bitrates for speech signals. Such studies all converge on the competitive qualitative performance as compared to alternative voice-specialized codecs~\cite{ramo2011voice,orosz2013performance,han2014determination}. Turning towards music signals, Hoene et al.~\cite{Hoene2011} published a foundational study where blind tests were performed using the MUSHRA methodology~\cite{mushra} on speech and music content with bitrates spanning 64 to 96~kbps, highlighting Opus' statistical superiority to alternatives such as HE-AAC, Vorbis or iLBC. In their respective studies, Jokisch et al.~\cite{jokisch2016audio}, Maruschke et al.~\cite{maruschke2016quality} and Valin et al.~\cite{valin2009high} also showed the perceptual competitiveness of the Opus codec in the presence of vocal and music signals. Moreover, thorough independent testing lead by the Hydrogenaudio community~\cite{hydrogenaudio} led to assess the superiority of the Opus codec in the presence of music~\cite{itu-r}. The authors employed the ABC/HR methodology~\cite{itu-r}, arguing that it is the most suitable framework for evaluating minor impairments—a rationale we share. Although this last study hints that transparency to uncompressed format may be reached at higher Opus bitrates, it does not produce formal, peer-reviewed and reproducible results. To our knowledge, our study is the first attempt at formally assessing the perceptual transparency of high bitrate Opus for music content.

\section{Perceptual test design}
\label{sec:methodology}

We argue that MUSHRA~\cite{mushra} lacks the resolution for high-bitrate evaluation, as its heavily degraded anchors, included in all trials, induce range-equalizing biases and ceiling effects~\cite{zielinski2008some}. Consequently, we adopt the ITU-R BS.1116-3 (ABC/HR) double-blind, triple-stimulus methodology~\cite{itu-r}, which avoids context bias and provides the psychometric sensitivity necessary to evaluate perceptual transparency thresholds.

\textbf{Trials} The test follows the double-blind triple stimulus with hidden reference, as specified by~\cite{itu-r}. Each trial compares a disclosed uncompressed reference (Ref) against two randomized, synchronized blind excerpts (A and B) for the hidden reference and the compressed audio. Utilizing a customized GoListen interface~\footnote{https://github.com/QxLabIreland/listening-test} with keyboard shortcuts for seamless switching, subjects score the blind tracks on a one-decimal ITU-R scale from 1 (Very annoying) to 5 (Imperceptible)~\cite{thiede2000peaq}.

As per the methodology, listening sessions are organized in 2 phases: training sessions, then grading sessions.

\textbf{Training} Guided training sessions (2 to 5 subjects) aim to familiarize subjects with the experimental protocol and a representative subset of audio material, presented in increasing difficulty. Subjects are invited to discuss perceptual impairments, while keeping numerical evaluations private~\cite{itu-r}. During training sessions, authors detail the sensory and affective anchors of the grading scale and encourage subjects to always assign exactly 5.0 to one blind stimulus and a lower score to the other, even if requiring a best-guess, to prevent range-shifting~\cite{zielinski2008some}. Training sessions utilize an identical interface to the grading sessions, augmented with a "reveal response" button for immediate feedback.

\textbf{Grading} For each subject, grading sessions are organized within 48 hours of training and capped at 30 minutes (15–16 trials) with a one-hour minimum inter-session rest to prevent auditory fatigue. To mitigate presentation bias~\cite{itu-r}, subjects evaluate $M$ randomly selected trials for each of the $K$ tested bitrates. Crucially, every session included 32~kbps anchors to establish baseline listener expertise prior to analysis.

\textbf{Listening setup} Grading sessions are organized in isolated acoustic booths using Beyerdynamic DT 770 Pro (80 Ohm) closed-back headphones to meet ITU-R acoustic recommendations~\cite{itu-r}. The measured room's ambient noise ($<$~32~dBA) combined with the headphones' 18 dBA passive isolation ensure a noise floor at the listener's ear well below the ITU-mandated NR 15 threshold, preventing artifact masking~\cite{itu-r}. Finally, hardware-induced spectral coloration is eliminated using model-specific flat-frequency calibration profiles via the SoundID Reference software~\footnote{https://www.sonarworks.com/soundid-reference}.

\textbf{Audio dataset} Following recommendations to stress-test the codec, we assemble a 23-sample dataset comprising 7 worst-case samples from a reference Hydrogenaudio benchmark~\footnote{https://opus-codec.org/comparison/}~\footnote{https://listening-tests.hydrogenaudio.org/igorc/samples.html}, augmented with 16 curated excerpts spanning various musical genres and drawn from high-fidelity reference playlists specifically for their challenging acoustic properties. This collection intentionally disproportionately features dense harmonic structures, such as plucked string instruments, known to be challenging for the codec~\cite{maruschke2016quality,hydrogenaudio}. Encoding utilize libopus~1.6.1 in music mode (CELT) to bypass early-frame speech/music detection artifacts. Unconstrained VBR is activated, causing 17 of the 23 tracks to encode above their target bitrates, confirming their challenging difficulty to be encoded by Opus. Lastly, encoded and original tracks are all normalized to -23 LUFS (EBU R 128) to eliminate loudness preference bias~\cite{zielinski2008some, ebu2011loudness}.

% \begin{figure*}[t]
%     \centering
%     \includesvg[width=\linewidth]{figs/score_per_sample_part1}
%     \includesvg[width=\linewidth]{figs/score_per_sample_part2}
%     \caption{Gradings for each sample and bitrate condition in experimental phase 1.}
%     \label{fig:boxplot_phase1}
% \end{figure*}

\textbf{Listening panel} The panel comprises 26 volunteer employees from Deezer. A preliminary questionnaire established their backgrounds: 7 females, 19 males, aged 22–49 with a mean of 36.5, 18 musicians, 12 with audio manipulation experience, and 2 self-declared as expert listeners in subjective evaluation. Naturally, the authors are excluded from the panel.

\section{Experiments}
\label{sec:experiments}

\subsection{Two phases}

To capture both general codec performance and the precise threshold of perceptual transparency, the evaluation utilizes a sequential two-phase architecture. Phase 1 establishes a broad baseline by evaluating the full panel (26 subjects) and complete corpus (23 excerpts) at bitrates of 32~kbps, 64~kbps, and 128~kbps, with subjects completing up to 30 randomized trials. Leveraging these results, Phase 2 establishes a rigorous worst-case stress-test by restricting the evaluation to the 10 most critical listeners and the 8 excerpts most challenging for the Opus codec, i.e. the excerpts where compression artifacts prove easiest to hear. To pinpoint transparency under these constrained conditions, subjects complete 32 trials across an expanded bitrate range of 32~kbps, 96~kbps, 128~kbps, and 160~kbps.

\subsection{Diffgrades}
\label{sec:diffgrades}

As per the recommendation~\cite{itu-r}, we consider Difference Grades, or \emph{Diffgrades}, defined as the the compressed sample grade minus the reference sample grade. For bitrate $i \in [\![1,I]\!]$, subject $j \in [\![1,J]\!]$ and sample $k \in [\![1,K]\!]$, we denote the Diffgrade as $D_{i,j,k} \in [-4, 4]$.
Negative Diffgrades indicate correctly identified compression artifacts, while positive Diffgrades reflect incorrect penalization of the reference; therefore, true perceptual transparency yields a symmetrical distribution with an expected mean Diffgrade of $0$.

\subsection{Subject post-screening}
\label{sec:post-screening}

To ensure data integrity per~\cite{itu-r}, a two-stage post-screening is used to exclude invalid subjects. First, a compliance filter eliminate listeners who violate the forced-choice protocol by registering undecided ties (Diffgrade $0.0$) on over a third of their trials, preventing artificial bias toward transparency. Second, an auditory discrimination filter employs a one-sided t-test ($\alpha = 0.05$, $H_0: \mu \ge 0$) on the 32~kbps anchor to exclude subjects unable to reliably distinguish severe degradation. Only subjects who pass both tests are retained for the statistical analysis.

\subsection{Linear mixed model}
\label{sec:lmm}

Observations from both phases are analyzed using a Linear Mixed-Effects Model (LMM) to robustly accommodate the sparse, randomized experimental design. Unlike traditional ANOVAs or t-tests, which mishandle missing observations and inflate Type 1 errors, the LMM specifies bitrate as a \emph{fixed effect} with \emph{crossed random intercepts} for subjects and samples. This architecture mathematically decouples true perceptual degradation from listener sensitivity and stimulus discriminability, yielding unbiased estimates of perceptual transparency. Formally, the statistical test evaluates the fixed effect coefficients ($\beta_i$) independently for each bitrate level $i$. The null hypothesis ($H_0^i$) postulates that a specific bitrate condition is perceptually indistinguishable from the uncompressed reference, yielding a population mean Diffgrade of zero. The alternative hypothesis ($H_1^i$) postulates that the compression introduces perceptible artifacts, resulting in systematic degradation:
$$H_0^i: \beta_i = 0 \text{ for bitrate }i$$
$$H_1^i: \beta_i < 0 \text{ for bitrate }i$$

To estimate the absolute mean Diffgrade for each operational bitrate, the model was specified using a cell-means coding structure. By suppressing the global intercept, it calculates an independent fixed-effect coefficient for every bitrate within a single framework. Formally,

$$D_{i,j,k} = \beta_{i} + u_j + v_k + \epsilon_{i,j,k}$$

where $\beta_i$ represent the fixed-effect parameter estimates for each tested bitrate $i$; $u_j$ is the random intercept associated with subject $j$, representing individual assessor criticality with $u_j \sim \mathcal{N}(0, \sigma_{\text{subject}}^2)$; $v_k$ is the random intercept associated with sample $k$, representing stimulus-specific encoding difficulty with $v_k \sim \mathcal{N}(0, \sigma_{\text{track}}^2)$; and $\epsilon_{i,j,k}$ represents the residual trial-level stochastic error with $\epsilon_{i,j,k} \sim \mathcal{N}(0, \sigma_{\text{residual}}^2)$. All random components $u_j$, $v_k$ and $\epsilon_{i,j,k}$ are assumed to be mutually independent.

\begin{figure*}[t]
    \centering
    \def\svgwidth{0.9\linewidth}
    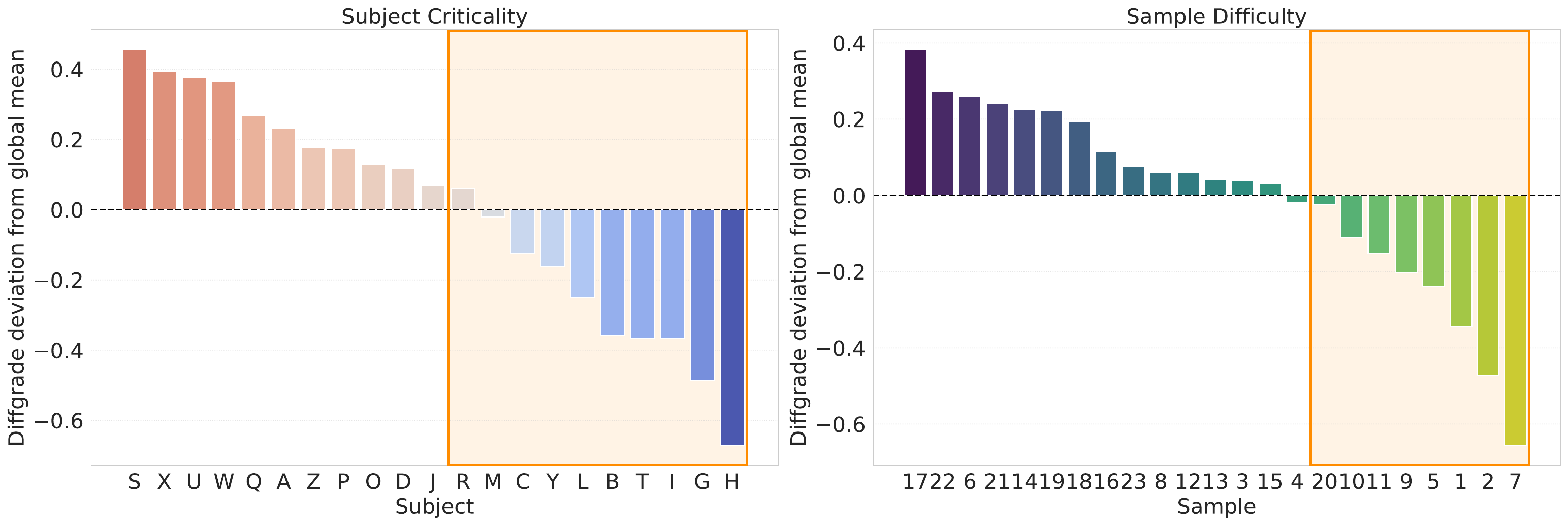
    % \includesvg[width=0.9\linewidth]{figs/blups_criticality_difficulty}
    \caption{BLUPs from subject random effect (left) from least to most critical, and sample random effect (right) from least to most audible artifacts. Highlighted regions show the selected candidates for Phase 2 experiments.}
    \label{fig:blups}
\end{figure*}

\section{Results}
\label{sec:results}

\begin{table}
\centering
\begin{tabular}{l|llll}
\toprule
    & Bitrate & Mean $\beta_i$ & 95\% confidence & p-value \\
\midrule
    \multirow{3}{*}{\rotatebox{90}{Phase 1}} & 32 & -1.66 & [-1.88, -1.45] & $< 10^{-3}$ \\
    & 64 & -0.44 & [-0.65, -0.22] & $< 10^{-3}$ \\
    & 128 & -0.08 & [-0.30, 0.13] & $4.56 \times 10^{-1}$ \\
\midrule
    \multirow{4}{*}{\rotatebox{90}{Phase 2}} & 32 & -2.98 & [-3.20, -2.77] & $< 10^{-3}$ \\
    & 96 & -0.22 & [-0.43, -0.004] & $4.54 \times 10^{-2}$ \\
    & 128 & -0.24 & [-0.45, -0.02] & $2.97 \times 10^{-2}$ \\
    & 160 & -0.04 & [-0.25, 0.18] & $7.47 \times 10^{-1}$ \\
\bottomrule
\end{tabular}
\caption{LMM analysis for Phases 1 and 2}
\label{tab:stats1} 
\end{table}

% \begin{figure*}[t]
%     \centering
%     \includesvg[width=\linewidth]{figs/score_per_sample_w2}
%     \caption{Gradings for each sample and bitrate condition in experimental phase 2.}
%     \label{fig:boxplot_phase2}
% \end{figure*}

\hspace{\parindent}\textbf{Phase 1} Post-screening (Sec.~\ref{sec:post-screening}) eliminated 5 subjects (2 for protocol violations, 3 for low-anchor failure), leaving 21 participants. The LMM analysis (Tab.~\ref{tab:stats1}, top rows) demonstrates that 128~kbps does not produce statistically significant degradation, yielding a mean Diffgrade of $-0.082$ (95\% CI $[-0.298, 0.134]$, $p = 0.456$) that is indistinguishable from the reference. Conversely, lower bitrates show significant degradation: 64~kbps exhibits mild impairment (mean $-0.438$, 95\% CI $[-0.654, -0.222]$, $p < 0.001$), while 32~kbps demonstrates severe artifacts (mean $-1.661$, 95\% CI $[-1.877, -1.445]$, $p < 0.001$). Finally, the random effects structure validates the LMM architecture: both listener sensitivity ($\sigma_{\text{subject}} = 0.346$, SE = $0.059$) and track discriminability ($\sigma_{\text{track}} = 0.286$, SE = $0.042$) introduce substantial standard deviation. This confirms the LMM successfully controls for both dominant psychometric biases and acoustic stimulus variations to effectively isolate true codec capabilities.

\textbf{Phase 2} To identify the most critical subjects and artifact-revealing samples for the Phase 2 stress-test, we extract Best Linear Unbiased Predictors (BLUPs) from the LMM’s random effects structure~\cite{robinson1991blup}, presenting them in Fig.~\ref{fig:blups}. Unlike simple averages, which are biased by sparse randomized trials, BLUPs apply statistical shrinkage to down-weight extreme, variable scores. Leveraging these predictors, we engineer a worst-case scenario by isolating the 10 most critical expert listeners and the 8 most revealing audio samples to test the codec's absolute perceptual limits in Phase 2.

Phase 2's stress-test (Tab.~\ref{tab:stats1}, bottom rows) proves the necessity of worst-case evaluations. Under strict expert scrutiny, the previously transparent 128~kbps condition is conclusively detected as degraded (mean Diffgrade $-0.236$, 95\% CI $[-0.449, -0.023]$, $p = 0.030$). Lower bitrates suffer variable penalizations (96~kbps: $-0.217$, $p = 0.045$; 32~kbps: $-2.982$, $p < 0.001$), and 160~kbps is established as the minimum condition to achieve no significant distinguishability (mean $-0.035$, 95\% CI $[-0.248, 0.178]$, $p = 0.747$). Variance analysis validates the BLUP selection: subject standard deviation collapses ($\sigma_{\text{subject}} = 0.118$, SE = $0.020$) and stimulus variance is heavily reduced ($\sigma_{\text{track}} = 0.195$, SE = $0.040$), confirming a homogeneously strict panel and uniformly challenging acoustic stimuli under this setup, highly challenging to Opus. 

\textbf{Equivalence testing} To validate perceptual transparency beyond standard difference testing, we employ a Two One-Sided t-Tests (TOST) equivalence analysis~\cite{lakens2017equivalence}. To properly account for the clustered nature of the experimental design, TOST is applied directly to the fixed-effect estimates and standard errors derived from the LMM, avoiding overly punitive multiple-comparison corrections on an already conservative variance model. The equivalence margin setting ($\Delta$) follows ITU-R BS.1387 (PEAQ)~\cite{thiede2000peaq}, where $\Delta = \pm 0.25$ is defined as a minimum tolerance margin when reconciling objective algorithmic outputs with subjective human evaluations, acknowledging that grading fluctuations within this band represent the practical noise floor of human perceptual scaling. In Phase 1, while 128~kbps approximates transparency, its variance causes it to narrowly fail strict statistical equivalence ($p = 0.064$, 95\% CI $[-0.298, 0.134]$), alongside 32~kbps and 64~kbps which conclusively fail ($p \approx 1.0$). Under Phase 2's critical expert constraints, 128~kbps diverges further from equivalence ($p = 0.450$). Absolute transparency in this worst-case scenario is uniquely established at 160~kbps, which successfully passes statistical \emph{equivalence} ($p = 0.024$), additionally confirming no statistical \emph{difference} (0 included in the 95\% CI $[-0.248, 0.178]$). We conclude that while 128~kbps exhibits no significant degradation for general listeners, 160~kbps is required to statistically reach perceptual equivalence.

\begin{figure}
    \centering
    \def\svgwidth{\linewidth}
    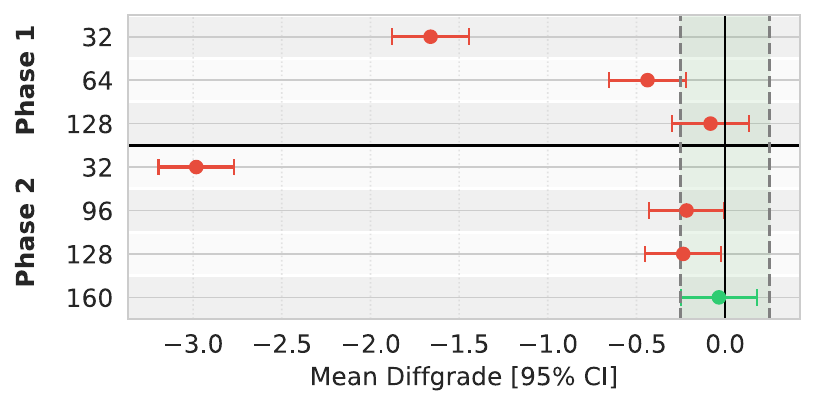
    % \includesvg[width=\linewidth]{figs/tost_forest_combined}
    \caption{TOST equivalence forest plot of Diffgrades for all bitrate conditions and $\Delta = \pm 0.25$. Green intervals are conditions statistically equivalent to uncompressed quality.}
    \label{fig:forest}
\end{figure}

\section{Conclusion}

This study determined the absolute perceptual transparency threshold for Opus audio using a rigorous two-phase ABC/HR methodology. To overcome limitations of aggregated statistics, a LMM was deployed to isolate true codec degradation from subject and sample variances. Phase 1 (21 listeners, 23 excerpts) established that while 128~kbps exhibits no statistically significant degradation for a general audience ($p = 0.456$), its variance causes it to narrowly fail strict TOST equivalence ($p = 0.064$, $\Delta=\pm 0.25$), whereas 32~kbps and 64~kbps exhibited severe degradation. By extracting BLUPs, Phase 2 subjected the codec to an extreme stress-test using only the 10 most critical subjects and 8 most artifact-revealing samples. Under these constraints, 128~kbps degraded further and failed equivalence testing, while transparency was exclusively established at 160~kbps, which resisted expert detection ($p = 0.747$) and passed strict equivalence ($p = 0.024$, $\Delta=\pm 0.25$). 
Ultimately, these results demonstrate that Opus encoding at 160~kbps achieves transparent music reproduction relative to uncompressed references under these rigorous testing constraints, though further research involving varied audiophile equipment and broader expert subject sampling would be beneficial to fully guarantee perceptual transparency across all real-world acoustic scenarios.

\bibliographystyle{IEEEbib}
\bibliography{refs}

\begin{thebibliography}{10}

\bibitem{valin2012definition}
Jean-Marc Valin, Koen Vos, and Timothy Terriberry,
\newblock ``Definition of the opus audio codec,''
\newblock Tech. {R}ep. RFC-6716, 2012.

\bibitem{valin2016high}
Jean-Marc Valin, Gregory Maxwell, Timothy~B Terriberry, and Koen Vos,
\newblock ``High-quality, low-delay music coding in the opus codec,''
\newblock {\em arXiv preprint arXiv:1602.04845}, 2016.

\bibitem{ramo2011voice}
Anssi Ramo,
\newblock ``Voice quality characterization of ietf opus codec,''
\newblock in {\em Proc. Interspeech}, 2011.

\bibitem{orosz2013performance}
P{\'e}ter Orosz, Tam{\'a}s Skopk{\'o}, Zolt{\'a}n Nagy, and Tam{\'a}s Lukovics,
\newblock ``Performance analysis of the opus codec in voip environment using qoe evaluation,''
\newblock in {\em Proc. ICSNC}, 2013.

\bibitem{han2014determination}
Yi~Han, Damien Magoni, Patrick McDonagh, and Liam Murphy,
\newblock ``Determination of bit-rate adaptation thresholds for the opus codec for voip services,''
\newblock in {\em Proc. IEEE ISCC}. IEEE, 2014, pp. 1--7.

\bibitem{Hoene2011}
Christian Hoene, Jean-Marc Valin, Koen Vos, and Jan Skoglund,
\newblock ``Summary of opus listening test results,'' 2011.

\bibitem{mushra}
ITU-R,
\newblock ``Method for the subjective assessment of intermediate quality level of coding systems,''
\newblock {\em ITU-R BS. 1534}, 2001.

\bibitem{jokisch2016audio}
Oliver Jokisch, Michael Maruschke, Martin Meszaros, and Viktor Iaroshenko,
\newblock ``Audio and speech quality survey of the opus codec in web real-time communication,''
\newblock in {\em Proc. ESSV}, 2016, vol.~3, pp. 254--62.

\bibitem{maruschke2016quality}
Michael Maruschke, Oliver Jokisch, Martin Meszaros, Franziska Trojahn, and Mario Hoffmann,
\newblock ``Quality assessment of two fullband audio codecs supporting real-time communication,''
\newblock in {\em Proc. ICSC}. Springer, 2016, pp. 571--579.

\bibitem{valin2009high}
Jean-Marc Valin, Timothy~B Terriberry, Christopher Montgomery, and Gregory Maxwell,
\newblock ``A high-quality speech and audio codec with less than 10-ms delay,''
\newblock {\em IEEE TASLP}, vol. 18, no. 1, pp. 58--67, 2009.

\bibitem{hydrogenaudio}
Hydrogenaudio community,
\newblock ``Opus music encoding quality,'' \url{https://wiki.hydrogenaudio.org/index.php?title=Opus}.

\bibitem{itu-r}
ITU-R,
\newblock ``Methods for the subjective assessment of small impairments in audio systems,''
\newblock {\em ITU-R BS. 1116-3}, 2015.

\bibitem{zielinski2008some}
Slawomir Zielinski, Francis Rumsey, and S{\o}ren Bech,
\newblock ``On some biases encountered in modern audio quality listening tests-a review,''
\newblock {\em JAES}, vol. 56, no. 6, pp. 427--451, 2008.

\bibitem{thiede2000peaq}
Thilo Thiede, William~C. Treurniet, Roland Bitto, Christian Schmidmer, Thomas Sporer, John~G. Beerends, and Catherine Colomes,
\newblock ``Peaq-the itu standard for objective measurement of perceived audio quality,''
\newblock {\em JAES}, vol. 48, no. 1/2, pp. 3--29, 2000.

\bibitem{ebu2011loudness}
EBU,
\newblock ``Loudness normalisation and permitted maximum level of audio signals,''
\newblock {\em EBU R 128-2010}, 2011.

\bibitem{robinson1991blup}
George~K Robinson,
\newblock ``That blup is a good thing: the estimation of random effects,''
\newblock {\em Statistical science}, pp. 15--32, 1991.

\bibitem{lakens2017equivalence}
Dani{\"e}l Lakens,
\newblock ``Equivalence tests: A practical primer for t tests, correlations, and meta-analyses,''
\newblock {\em Social Psychological and Personality Science}, vol. 8 (4), pp. 355--362, 2017.

\end{thebibliography}

\end{document}